# DeepGleason: a System for Automated Gleason Grading of Prostate Cancer using Deep Neural Networks

*Dominik Müller*[*,1,2], *Philip Meyer*[2], *Lukas Rentschler*[2,3], *Robin Manz*[2], *Jonas Bäcker*[2], *Samantha Cramer*[2], *Christoph Wengenmayr*[2], *Bruno Märkl*[3], *Ralf Huss*[3,4], *Iñaki Soto-Rey*[2], *Johannes Raffler*[2,5]

[1]IT-Infrastructure for Translational Medical Research, University of Augsburg, Germany

[2]Institute for Digital Medicine, University Hospital Augsburg, Germany

[3]Institute for Pathology and Molecular Diagnostics, University Hospital Augsburg, Germany

[4]BioM Biotech Cluster Development GmbH, Germany

[5]Bavarian Cancer Research Center (BZKF), Augsburg, Germany

**Abstract:**

*Background:* Prostate cancer, with over 68,000 annual diagnoses in Germany alone, is a widely prevalent and potentially lethal disease which necessitates efficient and reliable diagnostic tools to handle the growing demand for assessments based on standardized procedures like the Gleason scoring system. Advances in digital pathology and artificial intelligence (AI), particularly deep neural networks, offer promising opportunities for clinical decision support and enhancing diagnostic workflows. Previous studies already demonstrated AI's potential for automated Gleason grading, but lack state-of-the-art methodology, reusability, and model sustainability.

*Methods:* To address this issue, we propose DeepGleason: an open-source deep neural network based image classification system for automated Gleason grading using whole-slide histopathology images from prostate tissue sections. Implemented with the standardized AUCMEDI framework, our tool employs a tile-wise classification approach utilizing fine-tuned image preprocessing techniques in combination with a ConvNeXt architecture which was compared to various state-of-the-art architectures. The neural network model was trained and validated on an in-house dataset of 34,264 annotated tiles from 369 prostate carcinoma slides.

*Results:* In the performance evaluation, we demonstrated that DeepGleason is capable of highly accurate and reliable Gleason grading with a macro-averaged F1-score of 0.806, AUC of 0.991, and Accuracy of 0.974. The internal architecture comparison revealed that the ConvNeXt model was superior performance-wise on our dataset to established and other modern architectures like transformers. Furthermore, we were able to outperform the current state-of-the-art in tile-wise fine-classification with a sensitivity and specificity of 0.94 and 0.98 for benign vs malignant detection as well as of 0.91 and 0.75 for Gleason 3 vs Gleason 4 & 5 classification, respectively.

*Conclusions:* Our tool contributes to the wider adoption of AI-based Gleason grading within the research community and paves the way for broader clinical application of deep learning models in digital pathology. DeepGleason is open-source and publicly available for research application in the following Git repository: https://github.com/frankkramer-lab/DeepGleason.

## 1. INTRODUCTION

With an incidence of over 68,000 initial diagnoses per year in Germany alone, prostate cancer is a widely prevalent and potentially lethal disease having a considerable impact on global public health [1]. This substantial number of cases necessitates efficient and reliable diagnostic tools to handle the growing demand for accurate assessments. The diagnostic process for prostate cancer is highly standardized with the Gleason scoring system [2] being a pivotal component in determining the tumor's aggressiveness and, consequently, treatment decisions [3]. The reliable and robust identification of prostate cancer in histopathological tissue sections plays a crucial role in guiding clinical decision-making and treatment strategies.

In recent years, significant advances in digital pathology and artificial intelligence (AI) have opened up promising avenues for improving diagnostic robustness as well as efficiency [4]–[7]. Modern medical image analysis (MIA) models, particularly those based on deep neural networks, have demonstrated remarkable predictive capabilities in various medical applications [8], [9]. Notably, these models have shown the ability to automatically recognize and classify both well-established and novel patterns in digitized tissue sections [4]–[7]. Their potential in aiding the diagnosis of tumor diseases and predicting biomarkers has garnered considerable attention and interest within the medical community [8].

Automated deep neural network models can play a pivotal role in streamlining and enhancing the Gleason grading process, reducing the potential for interobserver variability and improving the overall consistency of diagnoses [10]. Recent works like Nagpal et al. [5] have showcased the potential of deep neural network based systems for Gleason grading of prostate carcinoma. These studies utilized vast datasets of over 1,500 annotated, digitized tissue sections in order to facilitate the training and evaluation of robust AI models [4], [5]. The successes of these studies have laid the groundwork for further advancements in AI-assisted prostate cancer diagnosis, opening up new opportunities for clinical application [11], [12].

Although researchers have successfully demonstrated the proof-of-concept application of AI algorithms for Gleason grading, they often have not published the deep learning model itself or a usable, ideally open-source, application. In light of the growing importance of AI-driven diagnostic tools in the medical field, clinical application of research findings is of critical significance. Additionally, it should be noted that these studies utilized neural network architectures from 2015 [5], and since then, there have been significant advancements in image classification algorithms and performances [13]–[15]. Incorporating state-of-the-art methods is necessary to build a usable Gleason grading software that can deliver reliable and accurate results, thereby maximizing its potential clinical impact in prostate cancer diagnostics.

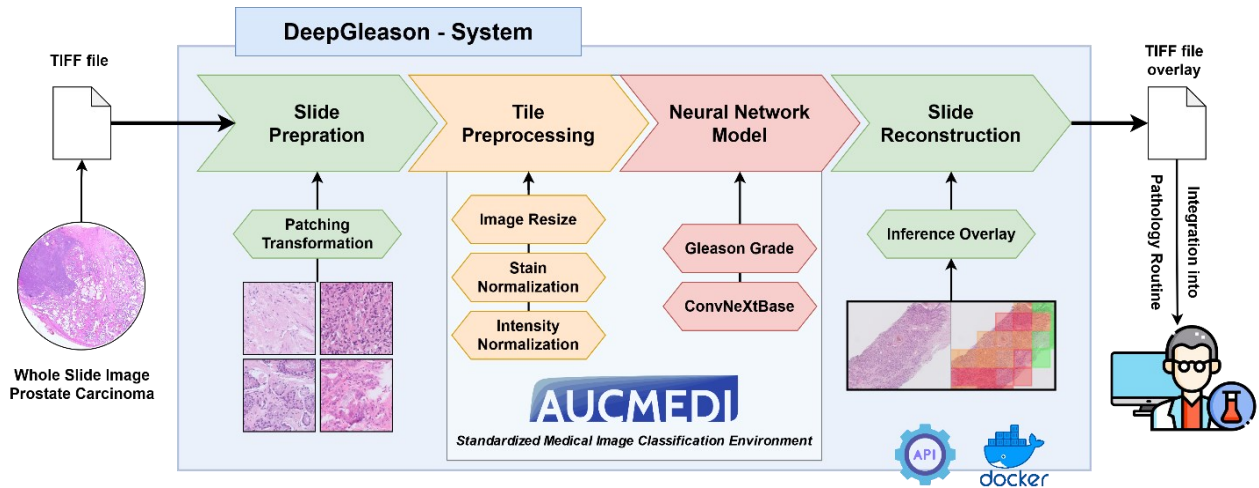

**Figure 1:** Workflow diagram of the DeepGleason system.

In this study, we develop an adaptive and open-source deep neural network based image classification system that can use whole-slide histopathology images (WSI) from prostate tissue sections to automatically classify the Gleason grade. Leveraging the latest advances in AI-based image analysis, we explore their potential for reliable classification of prostate cancer tissue sections, ultimately contributing to the ongoing efforts in improving patient outcomes, optimizing decision-making processes, and advancing the field of digital pathology.

## 2. MATERIAL AND METHODS

In this section, we present the methodology of our developed system, DeepGleason, designed for the classification of regions of interest (ROIs) in histological cross-sections based on the Gleason grading utilizing deep neural networks. The AI pipeline comprises four core modules: Slide preparation, tile preprocessing, the neural network model, and slide reconstruction. Figure 1 illustrates the workflow diagram of DeepGleason, providing a visual representation of the system's components and their interactions. The subsequent subchapters will delve into the detailed description of each core module, as well as outline the data acquisition procedure employed for model training and validation.

### 2.1. Dataset for Development and Validation

As retrospective data extraction, a total of 325 prostate carcinoma cases from 2019 to 2021 were initially identified in the laboratory information system of the Institute for Pathology and Molecular Diagnostics at the University Hospital Augsburg. A total of 1,202 slides of H&E-stained (hematoxylin- and eosin-stained) histological sections from prostate samples were associated to these cases, and corresponding Gleason grades were annotated for each slide based on available reports. 620 of these slides were digitized using a Philips UltraFast Scanner (UFS), extracted as iSyntax files from the digital image management system and converted to TIFF files utilizing the Philips Pathology SDK v2.0. From this pool, 369 digital slides were thoroughly annotated by a pathologist from the Institute for Pathology and Molecular Diagnostics at the University Hospital Augsburg. This annotation process involved marking specific image regions using polygons and classifying them into distinct categories, including regular tissue, carcinoma tissue based on Gleason grades 3, 4, or 5, artefacts such as air pockets, tissue distortion or slide contamination caused by sponges, and questionable regions that could not be definitively assigned to a specific Gleason class.

### 2.1.1. Data Preparation and Sampling

The annotated high-resolution digitized slides were preprocessed for the development process of our deep neural network model. For the digital slide preparation, we heavily utilized the VIPS package [16], which is an open image processing library that offers implementations for tile loading and lazy image processing pipelines. Thereby, each slide image was divided via patching into tiles with a size of 1024×1024 pixels (corresponds to 256×256 μm on the slide).

The labels provided by the pathologist were encoded as polygons used to mark regions of interest (ROIs) for each class. In order to calculate the annotated ROI coverage for a tile, the sum of the polygons was determined using grid-aligned bounding boxes, taking into account both the global coordinate minimum and maximum of each ROI. Since these ROIs could be either convex or non-convex polygons, special consideration was given to triangulations. This procedure enabled the representation of superordinate as well as fine-detailed spatial structures within the corresponding tiles, capturing a comprehensive view of the annotated features. Each tile was assigned a class according to the annotated class coverage if more than half of the tile's area (50% coverage) was annotated with a single class. Only for the artefact classes, we defined a required coverage of over 90% for a tile to be classified as 'artefact'.

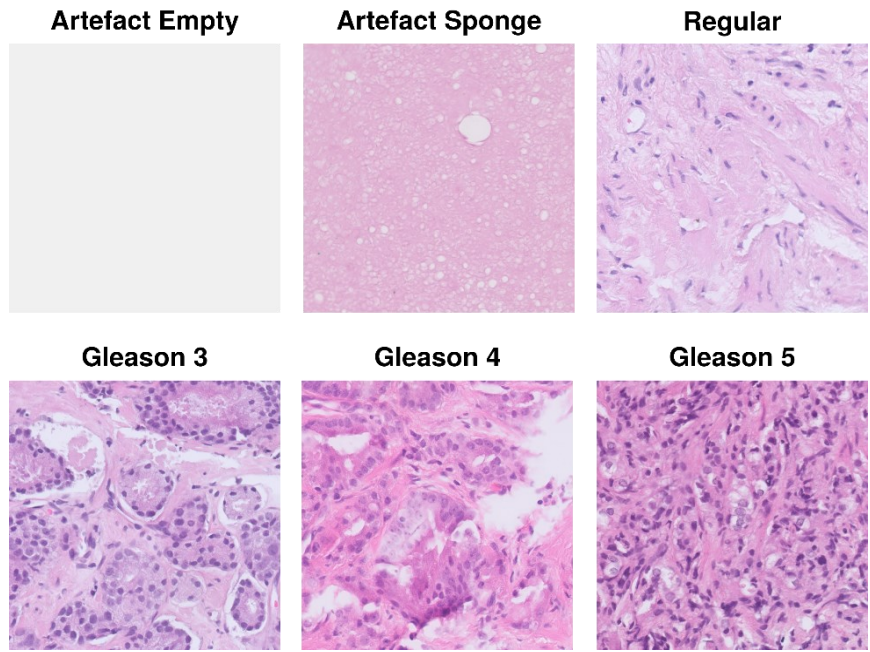

**Figure 2:** Visualization of tiles from prostate carcinoma histological sections of the training dataset.

As a result, it was possible to generate 78,564 tiles from the 369 slides. For reliable performance validation later, the set of tiles was sampled according to a 62-15-23 stratified percentage split resulting in 48,396 tiles for model training, 12,100 for validation, and 18,068 as a hold-out set for testing. Further frequency balancing for the Artefact Sponge class was conducted for the training and validation subset to obtain a corresponding class representation of 4%. As the Questionable region annotations had no use for model development and were evaluated in a separate study about analyzing explainable AI in pathology, we discarded all tiles with this annotation. This procedure reduced the number of tiles to 13,051 in the training, 3,264 in the validation, and 17,949 in the testing subset.

The class distribution revealed a significant imbalance, with a notable disparity in the number of tiles across classes: Regular - 7,916 (23.1%); Gleason 3 – 984 (2.9%); Gleason 4 - 2,878 (8.4%); Gleason 5 - 2,567 (7.5%); Artefact Empty - 4,487 (13.1%); Artefact Sponge - 15,432 (45.0). Samples of the corresponding classes are visualized in Figure 2.

### 2.2. Base Framework: AUCMEDI

Clinical application studies reveal significant challenges when integrating pipelines for image classification into a hospital environment [17]–[19]. Existing solutions are commonly developed outside clinical settings and optimized for individual datasets, leading to a lack of generalizability, thus rendering them impractical for reuse on other datasets and hindering their practical implementation in clinical research [19]–[21]. The open-source Python framework AUCMEDI [22], [23] provides a solution to these issues. This software package offers an API library for constructing standardized state-of-the-art medical image classification pipelines [22]. As a result, the framework enables the effortless establishment of a comprehensive as well as easily integrable pipeline for medical image classification and enables its practical deployment in clinical settings. Thus, our pipeline heavily utilized the AUCMEDI framework for image preprocessing and neural network model management, creating a standardized and robust medical image classification environment to ensure reproducibility in diverse clinical infrastructures.

### 2.3. Preprocessing and Image Augmentation

To improve the robustness and pattern-finding process, multiple preprocessing methods were exploited before passing the images to the neural network model.

For the training process, an extensive on-the-fly image augmentation procedure was applied including flipping, rotation, saturation, and hue augmentations. Other augmentation methods were excluded to avoid introducing artificial bias into the images. For preprocessing, the tiles were resized to 224×224 pixels according to the pretrained neural network architecture input shape. Subsequently, stain normalization based on Reinhard et al. [24] was applied in order to improve robustness between dissimilar digital slide scanners. Lastly, the pixel value intensities of the RGB tiles were normalized via the Z-Score normalization based on the ImageNet [25] mean and standard deviation.

### 2.4. Neural Network Models

For the development of a reliable and high-performing computer vision model, a total of five different deep neural network architectures were validated by comparing their predictive abilities. The following architectures were utilized: DenseNet121 [26], ResNeXt101 [15], Xception [27], ViT short for Vision Transformer (variant B16) [14], and ConvNeXt (variant Base) [13].

For the training process of the five pipelines, the following state-of-the-art procedure was applied. Transfer learning fitting of the classification head was conducted for 10 epochs based on weights obtained from the ImageNet dataset [25] using the Adam optimization [28] with an initial learning rate of $1E^{-4}$. Afterward, a fine-tuning fitting was conducted on the complete architecture with a maximal training time of 1,000 epochs and a dynamic learning rate starting from $1E^{-5}$ to a maximum decrease of $1E^{-7}$ (decreasing factor of 0.1 after 5 epochs without improvement on the monitored validation loss). Moreover, an early stopping technique was applied that stopped the training process after 10 epochs without validation loss improvement. The training was performed with a batch size of 28 samples, utilized the traditional epoch definition, and applied the weighted Focal loss by Lin et al. [29].

### 2.5. Performance Evaluation

The performance measurement was conducted on the tile-level with the described hold-out set in order to gain a detailed understanding of the classifier's reliability. For assessing the model performance, we computed a range of metrics, including Accuracy, F1-score, AUC (Area Under the Receiver Operating Characteristic curve), Sensitivity, and Specificity. For a comprehensive understanding of the metrics used and the strategies for classifier performance estimation, we refer to the excellent review by Maier-Hein et al. [30].

### 2.6. Software Packaging and Applicability

Packaging together the described pipeline, we developed an automated and streamlined system to facilitate straightforward reusability and clinical application of deep neural network based inference for Gleason grading.

The input data for our system consists of H&E WSI slides of prostate carcinoma. To prepare the data for application, the described patching strategy, dividing the whole slides into 1024×1024 pixel tiles, and the corresponding image preprocessing is used. Based on the validation results, the best-performing deep neural network model was selected and incorporated into the final system. For postprocessing, the Gleason grade predictions generated for each individual tile were visualized as heatmaps overlaying the original image. These resulting tiles are subsequently combined back into a single WSI. This approach allows seamless integration into clinical workflows, as the AI-based Gleason grading can be easily sighted within common pathology viewer through the returned WSI.

To further enhance the system's adaptability and integration in sensitive clinical environments, we encapsulated the entire system within a Docker container. This containerization enables uncomplicated usage, ensuring smooth deployment and compatibility in various clinical settings.

### 2.7. Software Availability

The DeepGleason system is hosted, supported, and version-controlled in the Git repository platform GitHub. This offers utilizing platform-hosted DevOps workflows and a hub for package documentation, community contributions, bug reporting as well as feature requests.

The Git repository is available under the following link: https://github.com/frankkramer-lab/DeepGleason.

The source code is licensed under the open-source GNU General Public License Version 3 (GPL-3.0 License), which allows free usage and modification for anyone.

## 3. RESULTS

In this section, we present the results of our developed prostate carcinoma classification system based on the Gleason grade. All five models, employing distinct neural network architectures, were successfully trained, with an average fitting requirement of 24.2 epochs and training times ranging from 8.3 to 42.0 hours per model. The total inference process for a single whole slide image took approximately 42.0 minutes on average (median) while the model-based tile classification excluding the slide preparation and reconstruction phase took approximately 8.4 minutes on average (median). All computations were performed on a NVIDIA DGX A100 (SXM4) workstation with 4 A100 GPUs (40 GB video memory), 500 GB memory and an AMD EPYC 7742 CPU (64 core and 128 threads). For performance evaluation, the models were thoroughly evaluated on the hold-out testing set.

In the architecture comparison, the ConvNeXtBase demonstrated superior performance with a macro-averaged F1-score of 0.806, AUC of 0.991, and Accuracy of 0.974. In contrast to traditional architectures like DenseNet121, Xception, and ResNeXt101 which performed moderately, modern architecture like ViT and ConvNeXt revealed strong prediction capabilities.

**Table 1:** Achieved results of the proposed image classification for prostate cancer.

| | DenseNet121 | | | ResNeXt101 | | | Xception | | | ViT B16 | | | ConvNeXt-Base | | |
|---|---|---|---|---|---|---|---|---|---|---|---|---|---|---|---|
| **Class** | Acc | F1 | AUC | Acc | F1 | AUC | Acc | F1 | AUC | Acc | F1 | AUC | Acc | F1 | AUC |
| **Regular** | 0.986 | 0.924 | 0.997 | 0.984 | 0.913 | 0.997 | 0.984 | 0.908 | 0.996 | 0.990 | 0.945 | 0.999 | 0.991 | **0.949** | 0.999 |
| **Gleason 3** | 0.984 | 0.475 | 0.980 | 0.986 | 0.517 | 0.977 | 0.979 | 0.450 | 0.979 | 0.984 | 0.522 | 0.977 | 0.987 | **0.557** | 0.984 |
| **Gleason 4** | 0.977 | 0.651 | 0.987 | 0.979 | 0.688 | 0.990 | 0.979 | 0.678 | 0.989 | 0.982 | 0.727 | 0.989 | 0.983 | **0.748** | 0.992 |
| **Gleason 5** | 0.989 | 0.721 | 0.994 | 0.989 | 0.738 | 0.995 | 0.989 | 0.714 | 0.994 | 0.992 | 0.764 | 0.993 | 0.993 | **0.823** | 0.998 |
| **Artefact Empty** | 0.939 | 0.772 | 0.977 | 0.944 | 0.785 | 0.975 | 0.945 | 0.789 | 0.982 | 0.943 | 0.782 | 0.966 | 0.946 | **0.796** | 0.984 |
| **Artefact Sponge** | 0.936 | 0.955 | 0.989 | 0.943 | 0.959 | 0.988 | 0.944 | **0.960** | 0.991 | 0.942 | 0.959 | 0.982 | 0.944 | **0.960** | 0.991 |

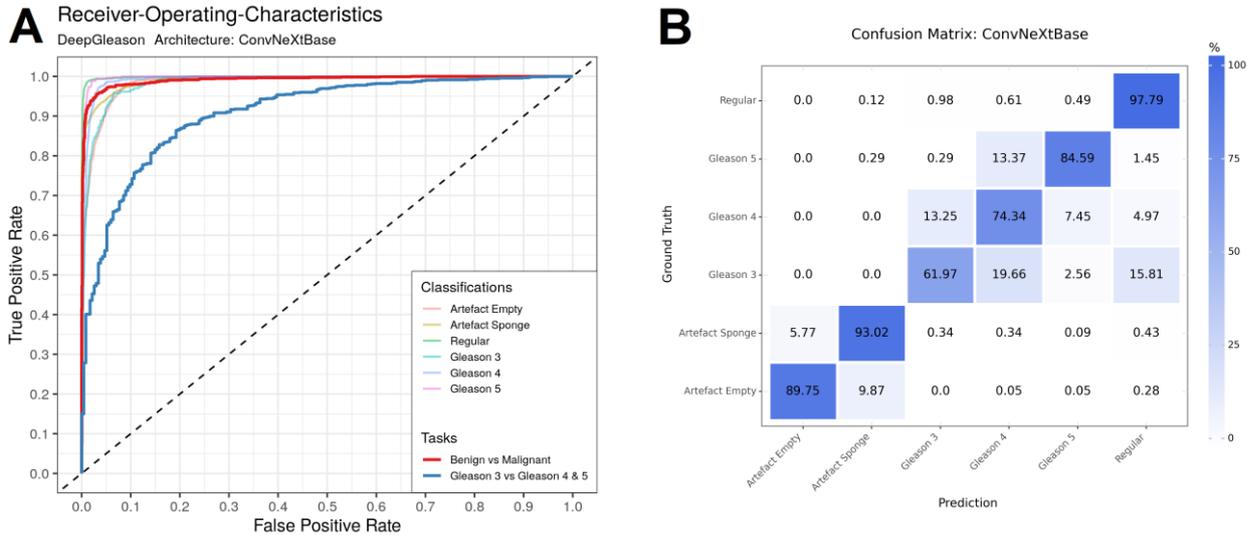

**Figure 3:** Achieved performance of the DeepGleason model – ConvNeXtBase.
**A:** Receiver operating characteristic curves for each class. **B:** Confusion matrix between prediction and annotation.

Still, traditional and modern architectures demonstrated both robust classification for Regular and Artefact Sponge tiles, in which the Xception architecture also obtained an F1-score of 0.960 for the Artefact Sponge class equal to the ConvNeXt. Nevertheless, the ViT B16 architecture was able to achieve the second-highest performance with only a marginal inferior macro-averaged F1-score of 0.783. A detailed overview of the achieved scores is shown in Table 1.

As illustrated in Figure 3, the ConvNext model mistook the 'Gleason 3' class, which also showed lower performance than the other classes, for 'Gleason 4' or 'Regular'. This scenario can be explained through the ordinal structure of the Gleason grading and revealed that the regions share high similarities with the adjacent Gleason classes representing the challenging task for reliable distinguishment.

## 4. DISCUSSION

The Gleason grading of prostate carcinoma represents a complex task within the medical domain due to the inherent intricacies associated with assessing histopathological patterns [3], [4], [10]. Prostate carcinomas manifest considerable histological heterogeneity, often characterized by the coexistence of various architectural patterns and cellular features [3], [10]. Discerning these subtle nuances is challenging, requiring a nuanced understanding of both normal and pathological tissue structures.

With DeepGleason, our system for automated Gleason grading based on deep neural networks, we were able to demonstrate a robust pipeline with remarkable performance which represents a noteworthy contribution to the domain of prostate cancer diagnosis. Our approach not only successfully reproduces the experiments by Nagpal et al. [5] but also surpassed them by incorporating state-of-the-art methods for enhanced accuracy and reliability. The high availability and usability of the pipeline contribute to its wider adoption within the research community and hold promise for broader clinical applications. The results of the class-wise evaluation revealed that our model is capable of providing reliable predictions even for challenging or discussion-worthy cases.

Deviations from the ground truth primarily involve adjacent classes, effectively capturing the ordinal structure of the Gleason classes. We were able to observe this in the confusion matrix (Figure 3), where the model faces particular difficulty in making precise differentiations between Regular, Gleason 3, and Gleason 4 tissue. The ordinal nature of Gleason grades demands meticulous expertise to distinguish between closely related grades. This challenge manifests also in our model predictions of individual classes, where subtle transitions between these classes pose the dominant hurdle for the model. Nevertheless, DeepGleason proved excellent capabilities for reliable Gleason grade approximation.

Another noteworthy observation emerged regarding the performance of newer architectures, such as ViT and ConvNeXt, in comparison to more established counterparts like DenseNet and ResNeXt. As expected, we were able to reproduce that advanced architectures presented stronger performance capabilities also in the domain of histopathology. This may be attributed to the increased model complexity and capacity allowing to capture intricate patterns in the training data more effectively, especially for differencing fine nuance between Gleason grades [13]. However, this enhanced capacity could lead to overfitting on insufficient-sized datasets, where the model becomes excessively tailored to the shades of the training set, hindering its ability to generalize to unseen data [31]. We assumed that this was the case for the ViT model as transformers commonly require large and diverse training data. In contrast, the ConvNeXt model exhibited to strike a balance between complexity and generalizability.

**Table 2:** Related work overview for deep learning based Gleason grading comparing model architecture and performance in cancer detection and fine classification (Acc: accuracy, Sens: sensitivity, Spec: specificity).

| Related Work | Model Architecture | Benign vs Malignant | | | Gleason 3 vs Gleason 4 & 5 | | |
|---|---|---|---|---|---|---|---|
| | | Acc. | Sens. | Spec. | Acc | Sens. | Spec. |
| Arvaniti et al. [34] (2018) | MobileNet | 0.82 | 0.85 | 0.79 | 0.77 | 0.81 | 0.74 |
| Bhattacharjee et al. [35] (2020) | Custom CNN | 0.94 | 0.92 | - | - | - | - |
| Duran-Lopez et al. [36] (2020) | Custom CNN (PROMETEO) | **0.98** | 0.98 | **0.98** | - | - | - |
| Karimi et al. [32] (2020) | MobileNet | 0.92 | 0.93 | 0.91 | 0.86 | 0.90 | 0.84 |
| Kott et al. [38] | ResNet | - | 0.93 | 0.90 | - | 0.83 | - |
| Nagpal et al. [5] (2019) | Xception | 0.81 | 0.81 | 0.79 | 0.76 | 0.77 | 0.74 |
| Marini et al. [39] (2021) | ResNet + DenseNet | - | - | - | 0.80 | 0.78 | 0.83 |
| Silva-Rodriguez et al. [37] (2021) | Custom CNN | 0.90 | 0.88 | 0.97 | 0.79 | 0.77 | **0.86** |
| Bulten et al. [10] (2022) | EfficientNet | - | **0.99** | 0.68 | - | - | - |
| Ikromjanov et al. [40] (2022) | Vision Transformer | 0.94 | 0.91 | 0.95 | 0.79 | 0.76 | **0.86** |
| Proposed: DeepGleason (2023) | AUCMEDI with ConvNeXt | 0.96 | 0.94 | **0.98** | **0.88** | **0.91** | 0.75 |

### 4.1. Comparison with Prior Work

For further evaluation, we compared our pipeline to other available Gleason grading approaches based on deep neural networks. Information and further details of related work were structured and summarized in Table 2. In order to enable increased comparability, we computed our performance on our testing dataset for cancer classification (benign vs malignant) and fine classification (Gleason 3 vs Gleason 4 & 5), which are popular task setups in the literature [10], [32].

In terms of performance, our proposed pipeline DeepGleason has achieved the highest specificity and ranking among the top-3 in sensitivity with 94% in distinguishing benign from malignant cases. This robust performance underscores its potential as a reliable tool for detection of malignant cases. For the fine classification between Gleason 3 and Gleason 4 & 5, DeepGleason achieved the highest accuracy and sensitivity, but with the trade-off of reduced specificity. This outcome can be attributed to the partwise smooth transition between Gleason 3 and Gleason 4, as also illustrated in the confusion matrix of Figure 3.

In terms of methodology, Table 2 indicates a preference for established and smaller deep neural network architectures. Surprisingly, even portable architectures like MobileNet [33], as demonstrated by Arvaniti et al. [34] and Karimi et al. [32], demonstrated robust predictive capabilities for both cancer detection and fine classification. Furthermore, it is noteworthy that custom CNN architectures designed specifically for Gleason grading continue to show promise besides established image classification architectures, as highlighted by the work of Bhattacharjee et al. [35], Duran-Lopez et al. [36], and Silva-Rodriguez et al. [37]. However, the predominant focus within the realm of deep neural networks remains on common image classification architectures such as ResNet and DenseNet [38], [39]. Other implementations of modern architectures like ViT, as presented by Ikromjanov et al. [40], tend to slightly lag behind in performance. The superior performance of smaller or less complex architectures could be attributed to their ability to recognize the broad cellular structures crucial for Gleason grading, particularly in routine cases. Nevertheless, it is essential to acknowledge that larger architectures may offer advantages, especially in handling more challenging cases, potentially providing valuable assistance to pathologists beyond routine scenarios. Still, another advantage of employing smaller architectures lies in their capabilities for more straightforward integration into clinical IT infrastructures through their reduced hardware requirements in terms of GPU and memory. This facilitates their practical implementation in real-world clinical settings and integration into pathology workflows, which is why we utilized the base variant of the ConvNeXt architecture instead of larger variants or transformers.

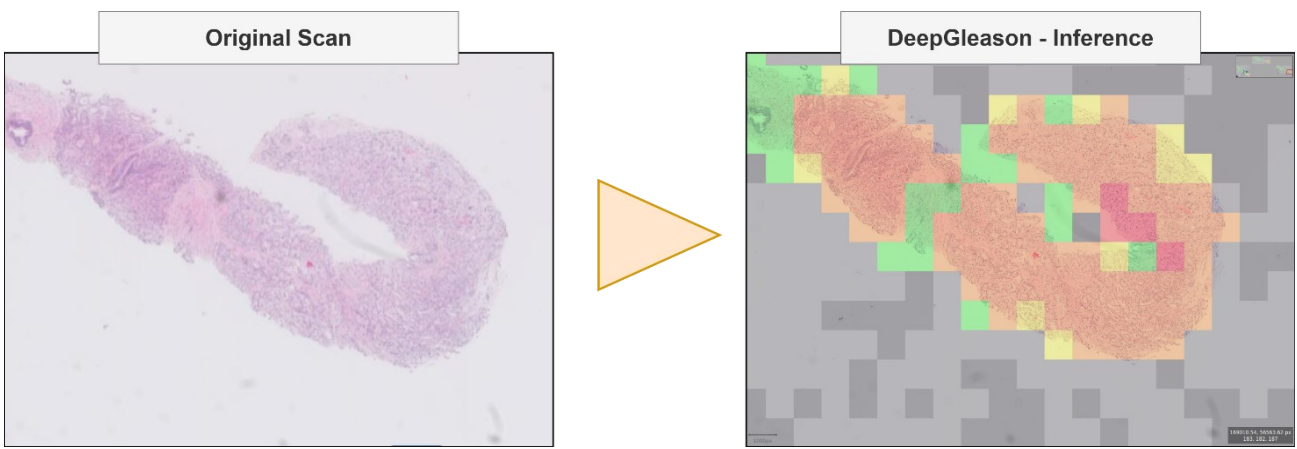

**Figure 4:** Visualization of the DeepGleason application on a H&E-stained prostate carcinoma whole-slide image scan. The coloring schema represents the following model classification: Green: Regular, Yellow: Gleason 3, Orange: Gleason 4, Red: Gleason 5.

Due to methodology similarities, the in-detail comparison with the work of Silva-Rodriguez et al. [37] is especially interesting. The authors reported F1-scores for their testing set of 0.86, 0.59, 0.54, and 0.61 for Regular, Gleason 3, Gleason 4, and Gleason 5 tiles, respectively [37]. In this comparison, DeepGleason demonstrated an overall superior Gleason grading of tiles apart from the Gleason 3 prediction. Nevertheless, in their confusion matrix for the tile-level classification on their testing set, Silva-Rodriguez et al. [37] also reported the challenging overlap between the Gleason 3 and Gleason 4 distinguishment with 228 Gleason 4 tiles incorrectly predicted as Gleason 3 (from 508 Gleason 3 predictions in total).

Compared to the work of Nagpal et al. [5], as mentioned in the introduction, we incorporated significant advancements in image preprocessing, image augmentation, state-of-the-art architectures, and the general pipeline setup to enhance reproducibility, applicability, as well as interoperability. These developments have enabled notable improvements in both cancer detection and fine classification performance. One of the major enhancements was the application of less aggressive image augmentation techniques to mitigate the potential introduction of biases. Additionally, different tile sizes were employed, with our study using tiles measuring 256×256 μm in contrast to their 911×911 μm tiles. Furthermore, we opted not to utilize ensemble learning, a multi-model strategy employed in the Nagpal et al. [5] study.

### 4.2. Limitations and Future Work

The Gleason grading system, as previously discussed, inherits notable subjectivity, relying on interpretations of glandular differentiation by pathologists [3], [10]. This complexity stemming from the subjective nature of the grading process underscores a key limitation in our current approach. The training data of our model pose a significant challenge as the model is exclusively trained on annotations from a single pathologist. Particularly in the transitions between classes, the encountered difficulties by the model could potentially be addressed by incorporating varying perspectives from multiple pathologists. This highlights our future work to enhance the model's performance by training on a consensus of multiple annotations for each slide. Such enhancement could also further improve our initial uncertainty detection mechanism which we plan to improve by integrating novel neural network uncertainty estimation methods from the literature [41], [42]. Despite these difficulties, the proposed system offers an excellent foundation for continuous improvement, especially in enhancing the fine-distinguishment of 'Gleason 3' regions. Future enhancements could involve integrating more detailed imaging data with a higher zoom factor, such as 128x128 μm or smaller tiles, to minimize information loss through resizing. Additionally, the implementation of ensemble learning offers strong potential, leveraging the strengths of multiple models trained on different deep learning architectures or disparate input image resolutions [43].

Another major opportunity for future work is the extension of Explainable Artificial Intelligence (XAI) [44] in DeepGleason. Through the AUCMEDI framework, our system already supports XAI methods like Grad-Cam [45], providing valuable insights into the otherwise opaque predictions of the AI model. Right now, we are investigating the usability and efficacy of XAI for Gleason grading in a comprehensive evaluation through qualitative and quantitative assessments involving multiple pathologists in an in-house clinical study. The results of this evaluation will be published in the near future, shedding light on the practical implications and benefits of incorporating XAI into the Gleason grading workflow.

## 5. CONCLUSIONS

In this paper, we developed and evaluated DeepGleason: a system for automated Gleason grading in whole slide images of prostate carcinoma using deep learning. Our tool employs a tile-wise classification approach utilizing fine-tuned image preprocessing techniques in combination with a ConvNeXt architecture. In the performance evaluation, we demonstrated that DeepGleason is capable of highly accurate and reliable Gleason grading while outperforming the current state-of-the-art in tile-wise fine-classification. DeepGleason is a free open-source tool whose availability and usability contribute to the wider adoption of AI-based Gleason grading within the research community and holds promise for broader clinical applications.

## DECLARATIONS

**Acknowledgements**
None.

**Conflicts of Interest**
The authors have no conflicts of interest to declare.

**Funding**
This work was supported in the context of the EKIPRO project funded by the Intramural Research and Young Scientist Grant from the Faculty of Medicine at the University of Augsburg.